\documentclass[a4paper]{article}

\usepackage{INTERSPEECH2020}
\usepackage{amsmath,cite}
\usepackage{url}

\expandafter\def\expandafter\UrlBreaks\expandafter{\UrlBreaks% save the current one
\do\a\do\b\do\c\do\d\do\e\do\f\do\g\do\h\do\i\do\j%
\do\k\do\l\do\m\do\n\do\o\do\p\do\q\do\r\do\s\do\t%
\do\u\do\v\do\w\do\x\do\y\do\z\do\A\do\B\do\C\do\D%
\do\E\do\F\do\G\do\H\do\I\do\J\do\K\do\L\do\M\do\N%
\do\O\do\P\do\Q\do\R\do\S\do\T\do\U\do\V\do\W\do\X%
\do\Y\do\Z\do\*\do\-\do\~\do\'\do\"\do\-}%
\usepackage{booktabs}

\DeclareMathOperator*{\argmax}{\arg\!\max}
\title{Peking Opera Synthesis via Duration Informed Attention Network}
\name{Yusong Wu\textsuperscript{1}\textsuperscript{$\ast$}\thanks{\textsuperscript{$\ast$}Yusong Wu performed the work while at Tencent.}, Shengchen Li\textsuperscript{1}, Chengzhu Yu\textsuperscript{2}, Heng Lu\textsuperscript{2}, Chao Weng\textsuperscript{2}, Liqiang Zhang\textsuperscript{2},  Dong Yu\textsuperscript{2}}
\address{\textsuperscript{1}Beijing University of Posts and Telecommunications \quad
\textsuperscript{2}Tencent AI Lab}

\email{\{wuyusong, shengchen.li\}@bupt.edu.cn, \\
\{czyu, bearlu, cweng, tatelqzhang, dyu\}@tencent.com}

\begin{document}
\maketitle
\begin{abstract}

Peking Opera has been the most dominant form of Chinese performing art since around 200 years ago. A Peking Opera singer usually exhibits a very strong personal style via introducing improvisation and expressiveness on stage which leads the actual rhythm and pitch contour to deviate significantly from the original music score. This inconsistency poses a great challenge in Peking Opera singing voice synthesis from a music score. In this work, we propose to deal with this issue and synthesize expressive Peking Opera singing from the music score based on the Duration Informed Attention Network (DurIAN) framework. To tackle the rhythm mismatch, Lagrange multiplier is used to find the optimal output phoneme duration sequence with the constraint of the given note duration from music score. As for the pitch contour mismatch, instead of directly inferring from music score, we adopt a pseudo music score generated from the real singing and feed it as input during training. The experiments demonstrate that with the proposed system we can synthesize Peking Opera singing voice with high-quality timbre, pitch and expressiveness.

\end{abstract}

\section{Introduction}
\label{sec:intro}

Peking Opera, also known as Beijing Opera or Jingju, is Chinese traditional performing art which combines music, vocal performance, mime, dance and acrobatics.
Singing in Peking Opera has various styles, each widely different depending on different role type and music styles. Strong personal styles also make the actual singing can be different from the given music notes. Like a dialect to Mandarin, it even has its unique way of pronunciation. Moreover, melody in singing often consist of arias with variation of complex transitory and vibratos, which makes the singing very expressive and difficult to learn. Another difference from normal singing is the note length has a great variance, sometime very long note can appear (can be more than 10 seconds). All above factors makes it very challenging to modelling and generating Peking Opera singing comparing to normal singing.

Although there are few works focusing on the synthesis of Peking Opera, or more broadly, opera, the synthesis of singing voice has been researched since 1962 when Kelly and Lochbaum~\cite{lochbaum1962speech} used an acoustic tube model to synthesis singing voice with success. Recently, several works~\cite{blaauw2017neural,yi2019singing,kaewtip2019enhanced,wada2018sequential,chandna2019wgansing,Hono2018} use deep neural networks to synthesis singing voice which, known as parametric systems, process fundamental frequency (or pitch contour, $f_0$) and harmonics features (or timbre) separately. As a typical case among such systems, Neural Parametric Singing Synthesizer (NPSS)~\cite{blaauw2017neural} using a phoneme timing model, a pitch model and a timbre model each consist a set of neural networks to generate acoustic parameters of the singing. In NPSS, a Fitting Heuristic method is introduced to eliminate the mismatch between music note duration and the predicted phoneme duration. However, Fitting Heuristic method is totally rule based and it requires to locate the principal vowel before adjusting phoneme duration. This maybe acceptable in most English or Japanese singing cases, but can cause huge duration error when synthesizing Peking Opera. Different from normal speech or singing, in Peking Opera, one syllable can last very long time and contains a long sequence of phonemes, e.g. ``l-j-E-a-a-N". More importantly, one can't simply tell which phoneme amongst all these phonemes is the principle phone. There could be multiple equally important phonemes in Peking Opera singing.

To better synthesize the expressive Peking Opera, this paper proposes a Peking Opera singing synthesis system based on Duration Informed Attention Network (DurIAN)~\cite{yu2019durian}. The main contribution in this study lies in the two following points: 1) To tackle with rhythm mismatch between music note duration and the predicted phoneme duration, contextual based mixture density networks (MDN)~\cite{bishop1995neural} followed by a Lagrange Multiplier optimization is proposed and implemented for duration modelling. This method is completely data-driven, and more importantly, skips the step of locating the principle phoneme from the conventional Fitting Heuristic method. 2) To deal with the melody mismatch between original music score and the actual singing, and also to better model the expressive variations and vibratos in Peking Opera, a pseudo music score is generated from the real singing and fed as input during DurIAN~\cite{yu2019durian} model training. Experimental Results show proposed duration modeling and prediction method outperforms the Fitting Heuristic method by a large margin. And the generated pitch contours also demonstrate our system's ability to synthesize the singing variations and vibratos in Peking Opera.

The following sections of this paper are organized as follows. Firstly, the proposed model architecture is introduced. Next, proposed Lagrange Multiplier-based duration prediction and pseudo score generation are introduced in Section 2. In section 3 experiments are conducted based on a unique Peking Opera database. Finally, a quick discussion and conclusion is given in Section 4.

\section{Methods}
\label{sec:method}

\begin{figure*}[bht] %figure*
 \centerline{
 \includegraphics[width=0.75\textwidth]{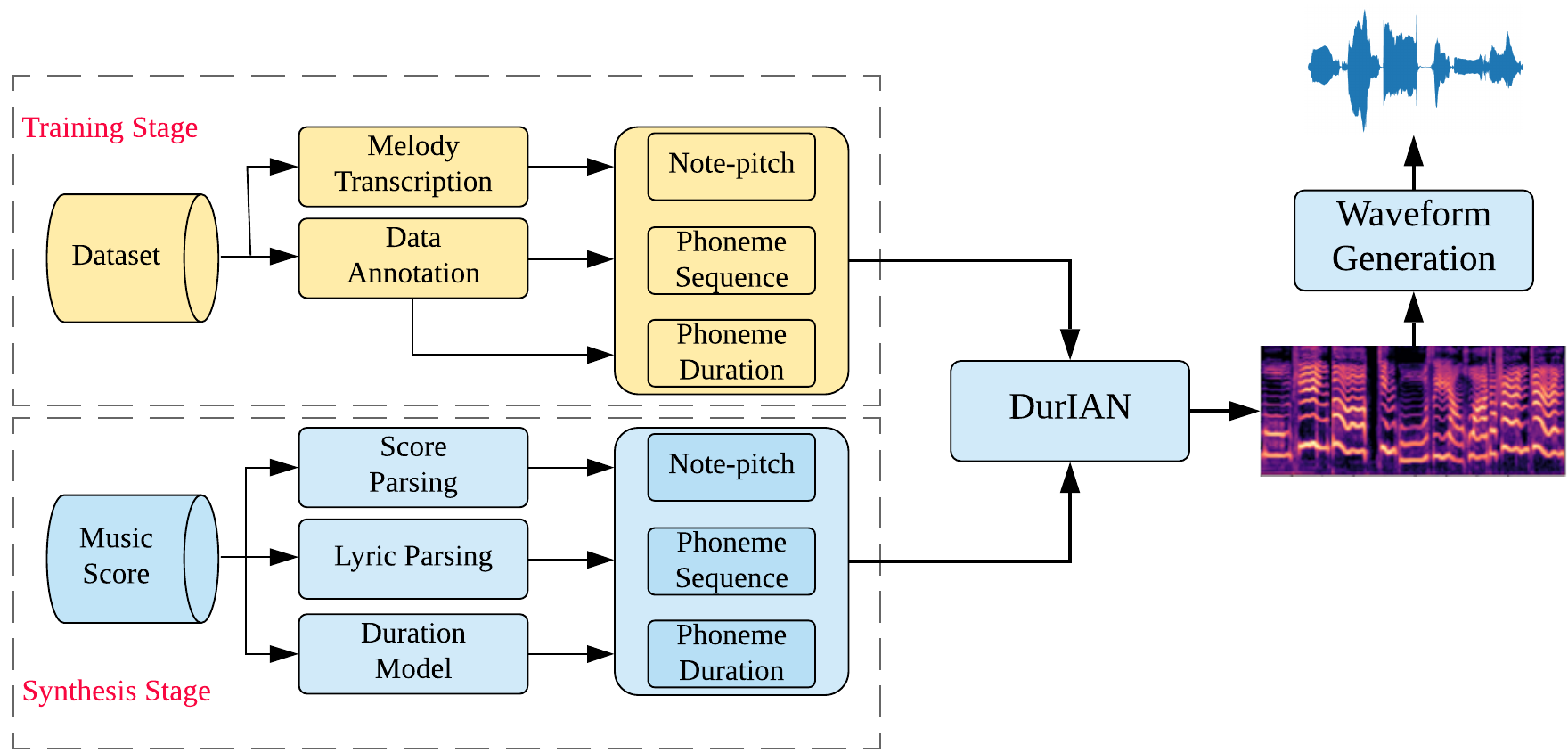}}
 \caption{The overview diagram of the proposed system. The proposed system is trained using data annotation and melody transcription result. In the synthesis stage, note-pitch and phoneme sequence are parsed from score, while the phoneme duration is predicted by the proposed duration model.}
 \label{fig:model}
 \vspace{-4mm}
\end{figure*}

In proposed system, three efforts made it possible to synthesize expressive Peking Opera singing from music score: 1) DurIAN~\cite{yu2019durian} based auto-regressive framework is used to generate output spectrogram features, with generated pitch as condition. 2) With the constraint of Lagrange Multiplier, a contextual dependent mixture density network (MDN) based phoneme duration modeling and generation method is introduced for accurate duration modelling. And 3) a melody transcription method is used to obtain a pseudo score from training waveform samples in training process. This step is to ensure the extracted acoustic feature and the input music score consistent in training as well as in inference. Fig.~\ref{fig:model} shows the training and synthesis diagram. In training stage, note-pitch of the singing is provided by melody transcription results while phoneme sequence and phoneme duration are obtained from data annotation. While in the synthesis, input note-pitch is from real music score and input phoneme sequence comes from lyric parsing module which also use music score as reference.

\subsection{Spectrogram Generation}

 Duration Informed Attention Network (DurIAN)~\cite{yu2019durian} structure is used here as main frame.
Similar to DurIAN, in proposed system, the features of temporal dependency in phoneme sequences is first modelled by a \textbf{phoneme encoder} consist of a two-layer prenet followed by a CBHG module~\cite{wang2017tacotron} (1-D convolution bank + highway network+ bidirectional GRU) for learning contextual dependencies of phonemes. The encoded phoneme sequence is aligned with output spectrogram in \textbf{alignment model} where singer identity and frame-wise generated pitch are further added. The final Mel-spectrogram frames are generated by a \textbf{decoder} with Gated Recurrent Unit (GRU)~\cite{cho2014learning} in an auto-regressive manner where the temporal dependencies of the Mel-spectrogram are modelled, and a post-net is used for refining the output. Different from DurIAN, a CBHG module is used as a post-net instead of a fully convolution network to improve generation quality.

In the training stage, phoneme sequences are obtained from data annotation. Ground-truth phoneme duration is used in phoneme sequence expanding in alignment model. The input music note is generated from real singing by the melody transcription module which is introduced in section 2.3. While in the synthesis stage, the phoneme duration is predicted by the trained duration model introduced in section 2.2 and the input music note is from the real music score.

\subsection{Phoneme Duration Modelling}
In the music score, the duration of each note and also the lyric aligned with note is given. However, the duration of each phoneme needs to be predicted in order to generate syllable or word pronunciation during synthesis stage. Thus, a phoneme duration model is trained in training stage and used to generate phoneme duration in generation process. Unlike in speech synthesis where no music score exists, in singing synthesis, the sum of predicted phonemes duration in a note should be always equal to the duration of the note. This is to make sure synthesized singing follows the rhythm on the score. To ensure the duration of the phoneme adds up to the note duration, Fitting Heuristic method is introduced in ~\cite{blaauw2017neural,blaauw2020sequence}.
However, Fitting Heuristic duration scaling relies heavily on the selection of primary vowel and $r_0$. In case of Peking Opera singing, there are prevailing use of very long notes. When scaling phoneme duration for a very long note, and sometimes the note can contain more than 5 equally important phonemes, the incorrect choice of primary vowel will lead to huge phoneme duration prediction errors. Instead, our system generates phoneme duration with Lagrange multiplier constraint which has been used in speech synthesis~\cite{yamagishi2006introduction,zen2007state,lu2009full}. All the scaling or compensate factors are generated through data-driven methods other than based on rules.

Bi-directional Long Short Term Memory (Bi-LSTM) networks~\cite{hochreiter1997long,schuster1997bidirectional} with mixture density output layer~\cite{bishop1995neural} is used as out phoneme duration model. With a total of $M$ phonemes in the given note from music score, the duration distribution $\mathcal{L}(d_i, \lambda)$ for each phoneme can be represented as

\vspace{-3mm}
\begin{equation}
   \mathcal{L}(d_i, \lambda) =  \sum_{k=1}^{K} \Pi_{i}^{k} N(\mu_i^{k},  (\sigma_i^{k})^2)
\end{equation}
and
\begin{equation}
\sum_{i=1}^{M} d_i = T,\ \ i = 1,2,...,M
\end{equation}

 where $K$ is the total number of Gaussian Mixtures for each phone. $\Pi_{i}^{k}$, $\mu_i^{k}$ and $\sigma_i^{k}$ are the weight, mean and standard deviation for the $k$th mixture of the $i$th phoneme respectively. $\lambda$ indicates the total MDN parameter set mentioned above. In training stage, the parameters in proposed Bi-LSTM based mixture density network duration model are learned through back-propagating the negative log-likelihood error.

 In synthesis stage, to maximize log MDN likelihood under the constraint of given total word duration, a Lagrange Multiplier is introduced and the task becomes to solve the following equation:

 \vspace{-3mm}
 \begin{equation}
 \label{E4}
    \nabla_{d_i, \alpha} [\sum_{i=1}^{M} log(\mathcal{L}(d_i, \lambda))-\alpha(\sum_{i=1}^{M}d_i-T)]=0
\end{equation}
 where $\alpha$ is the Lagrange Multiplier factor.

 Because there is no close form solution for the Multi-Gaussian distribution maximum likelihood optimization problems, to simply the form of final solution, we always choose the Gaussian with the maximum weight instead of using all mixtures when calculating. Thus,

 \vspace{-3mm}
\begin{equation}
   {d_i}\sim \mathcal{L}(d_i, \{{\mu_i^{k}, \sigma_i^{k}\}}) \quad \text{where} \quad k= \argmax_k \Pi_{i}^{k}
\end{equation}

Let $\hat{\mu_{i}}$, $\hat{\sigma_{i}}$  be the Gaussian means and standard deviations with the largest weight for the $i$th phone, and the solution for Eq.~\ref{E4} is formulated as:

\vspace{-3mm}
\begin{equation}
    d_{i}=\hat{\mu_{i}}+\hat{\sigma_{i}}^{2} \cdot \alpha \quad \text{where} \quad \alpha=\frac{T-\sum_{i=1}^{M}\hat{\mu_{i}}}{\sum_{i=1}^{M}\hat{\sigma_{i}}^2}
\end{equation}

As a comparison, if there's no Lagrange Multiplier constraint in Eq.~\ref{E4}, the result for the maximum likelihood will be only the means for the Gaussian $\hat{\mu_{i}}$.
The equation above can be interpreted as scaling of each phoneme duration prediction $\hat{\mu_{i}}$ by a compensation term $\alpha$ which is decided by 1) the gap between music score note duration and sum of predicted phoneme duration, and 2) its standard deviation. And that indicates phonemes with large variance, such as some vowels, can be extend or compress more than the ones with smaller variance. This result is actually consistent with the duration scaling rules in Fitting Heuristic, but more generally formulated, and obtained through purely data-driven method.

\subsection{Melody Transcription}

\begin{figure}[bht] %figure*
 \centerline{
 \includegraphics[width=0.8\columnwidth]{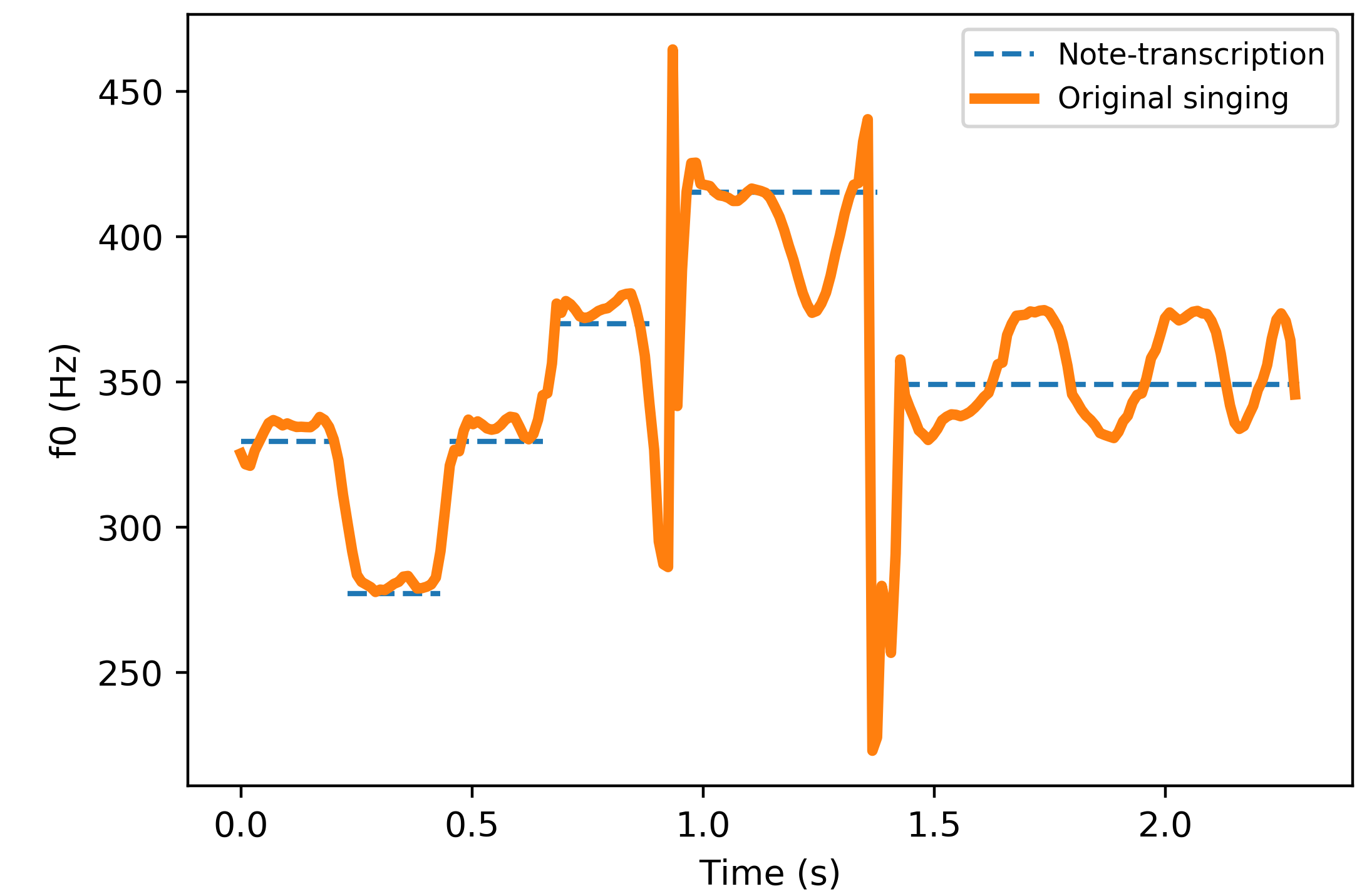}}
 \caption{A example of the melody transcription results on Peking Opera singing.}
 \label{fig:transcription-result}
\end{figure}

Although published version of the Peking Opera score is available, the improvisation and expressiveness of the singer makes the actual singing inconsistent with the score. Efforts have been made to build Peking Opera singing synthesis system directly from original music scores, but the model turned out to be badly trained, and the generation results were with poor quality. Alternatively, a pseudo music score is automatically transcribed from actual Peking Opera singing to replace original music score in training stage. Same melody transcription method proposed in~\cite{gong2016pitch} is used here. The melody transcription method is based on a Hidden Markov Model~\cite{mauch2015computer} based pitch tracking module which uses probabilistic YIN~\cite{mauch2014pyin} pitch estimation. First, frame-wise fundamental frequency of the singing with temporal smoothness is estimated by probabilistic YIN~\cite{mauch2014pyin} pitch estimation where a Hidden Markov Model is applied to decode from multiple pitch candidates. Then, the estimated pitch track is fed into another Hidden Markov Model to render frame-wise discrete note pitch. The Hidden Markov Model used in transcription process contains 3-state consisting of Attack, Stable and Silent. When transitioning from current note to the next note, probabilities of note transition are calculated by a note transition probability function. A Peking Opera genre-specific note transition distribution are used for better note pitch decoding, achieving a note transcription F-score of 0.73. The output of the transcription algorithm is frame-wise discrete note pitch of the singing, where possible pitch value ranges from MIDI pitch 35 (B1) to MIDI pitch 85 (C\#6) with 3 steps per semitone.
Here, note pitch output are further quantified to integral MIDI pitch, and pitch of the unvoiced frame is set to 0. An example of the transcribed melody is shown in Fig.\ref{fig:transcription-result}, where the orange line shows the $f_0$ of the singing pitch and the blue dotted line indicates the transcribed result.

\section{Experiments and results}
\label{sec:experiment}

To demonstrate the effectiveness of proposed methods, two experiments are conducted. In phoneme duration prediction experiment, the objective predicted phoneme duration error is compared between proposed duration model and Fitting Heuristic method. In Peking Opera synthesis experiment, generated pitch contours are drawn and subjective Mean Opinion Score (MOS) test show our system is capable of synthesizing Peking Opera with fair quality.

\subsection{Data}
\label{sec:data}

Although being popular for centuries, Peking Opera has received little effort in research field. To collect Peking Opera training data is not a trivial task. One of the available data set is the ``Jingju a cappella singing"  ~\cite{Gong2019,Gong2018,Gong2018b,Gong2018a}. ``Jingju a cappella singing" contains over 10 hours of a cappella Peking Opera recordings and annotations, but amongst which only 2 hours of recordings have been phoneme annotated.
The 2 hours of recordings with phoneme annotation contains 71 Peking Opera singing fragments. And the annotation of phoneme adopts a modified X-SAMPA (Extended Speech Assessment Methods Phonetic Alphabet) phoneme set containing 51 phonemes. The singing data is further segmented according to line boundary, resulting a set of 606 singing phrases. In the experiment part, all experiments are conducted using the 2 hours of the annotated dataset.

\subsection{Phoneme Duration Prediction}
In order to compare our Lagrange Multiplier-based phoneme duration prediction method with Fitting Heuristic based methods, six phoneme duration prediction models are implemented and compared. First, the duration model used in NPSS system~\cite{blaauw2017neural} is implemented (\textbf{CNN-Softmax-Fit}) which consist of 1d-CNNs and a softmax output predicting duration which are discretized to 50 log scale bins. To compare different duration scaling method, the softmax output in \textbf{CNN-Softmax-Fit} is then replaced with mixture density output with Lagrange Multiplier for duration optimization (\textbf{CNN-MDN-Lag}). Next, the duration model proposed in this paper is built (\textbf{BiLSTM-MDN-Lag}) which used a BiLSTM with mixture density output and Lagrange Multiplier optimization. For comparison, the output layer of the proposed duration model is replaced with a scalar prediction and Fitting Heuristic scaling while trained using mean square error (MSE) criterion (\textbf{BiLSTM-MSE-Fit}). To further compare the duration scaling method, the proposed system using Fitting Heuristic scaling instead of the Lagrange Multiplier optimization is built (\textbf{BiLSTM-MDN-Fit}) which use the mean of the predicted distribution as duration prediction.
Last, the duration prediction method originally used in~\cite{blaauw2020sequence} is implemented (\textbf{Mean-Fit}) which use the mean duration of each phoneme as a look-up table for duration prediction and Fitting Heuristic for scaling.

In the experiment, the frame length is set to 10ms. 64 Peking Opera singing phrases are randomly chosen from database for test, while the rest of the data is used for training. The CNN blocks are built according to~\cite{blaauw2017neural}. All Bidirectional-LSTMs use 2 hidden layers and has a 256-dimensional hidden layer size with training dropout rate of 0.5. And all Fitting Heuristic scaling set the second phoneme as the primary vowel. The number of Gaussian mixtures predict by model is set by 2 which we find yields best results.

The results are shown in Tab.~\ref{duration-results} with average duration prediction error per phone, in terms of number frames. For in Peking Opera singing, there could be extremely long notes which may last for seconds. Thus when counting the results, two kind of results are counted: 1) the music score notes shorter than 2 seconds case, and 2) all notes cases. This objective average phoneme prediction error results reveal the proposed duration model outperforms other methods, achieving minimum prediction error. Probably because the dynamic range of phoneme duration in Peking Opera is much larger than normal speech or singing, 50 discrete log scale bins is too coarse as the prediction target and thus introduces largest prediction error in \textbf{CNN-Softmax-Fit}.  From the results, we can see propose Lagrange Multiplier-based phoneme duration scaling consistently outperforms the Fitting Heuristic scaling by a large margin. And our proposed Mixture Density Network based Lagrange Multiplier phone duration generation method \textbf{BiLSTM-MDN-Lag} renders the best performance. It is worth noting that \textbf{BiLSTM-MDN-Fit} outperforms \textbf{BiLSTM-MSE-Fit} in predicting phoneme duration for shorter notes while perform worse when including long notes. Further analysis shows this happens when the music note is long, and when Fitting Heuristic scaling is employed, the generated phoneme duration is dominant by the stretching length. Moreover, when the principle vowel is locating wrong, the prediction error can be huge.

\begin{table}[]
\caption{The mean phoneme duration prediction error in number of frames for different duration model.}
\vspace{-4mm}
\label{duration-results}
\begin{center}
\begin{tabular}{@{}ccc@{}}
\toprule
                   & all           & notes \textless~ 2s \\ \midrule
BiLSTM-MDN-Lag (proposed) & \textbf{8.91} & \textbf{3.24}      \\
BiLSTM-MSE-Fit          & 13.63         & 4.23               \\
BiLSTM-MDN-Fit        & 17.15         & 4.16               \\
CNN-MDN-Lag             & 14.74          & 4.75             \\
CNN-Softmax-Fit~\cite{blaauw2017neural}     & 19.61     &5.83       \\
Mean-Fit~\cite{blaauw2020sequence}             & 18.69         & 5.52               \\ \bottomrule
\end{tabular}
\end{center}
\vspace{-8mm}
\end{table}

\subsection{Peking Opera Synthesis}
\subsubsection{Subjective MOS score}

In order to evaluate the subjective naturalness of the proposed Peking Opera synthesis system, Mean Opinion Score test is conducted where test participants are asked to score the singing samples from 1-5 according to the naturalness of the singing. 1 stands for most unnatural and unintelligible, 5 stands for most natural and almost the same as sung by real people. 2,3,4 are somewhere between. Peking Opera music scores in the format of MusicXML are used as score input which first parsed into notes and corresponding phonemes then input to the duration model to get each duration of the phonemes. Using the predicted phoneme duration along with notes and phonemes as input, singing samples are generated by the proposed system, and MOS test is conducted using the generated singing. In result, a 3.34 of MOS score is obtained, which indicates that the proposed system can generate Peking Opera singing with fair quality. Generated audio samples can be found on the web page: https://lukewys.github.io/files/Peking-Opera-Synthesis-2020.html.

\subsubsection{Generated Pitch Contour}

\begin{figure}[bht] %figure*
 \centerline{
 \includegraphics[width=0.8\columnwidth]{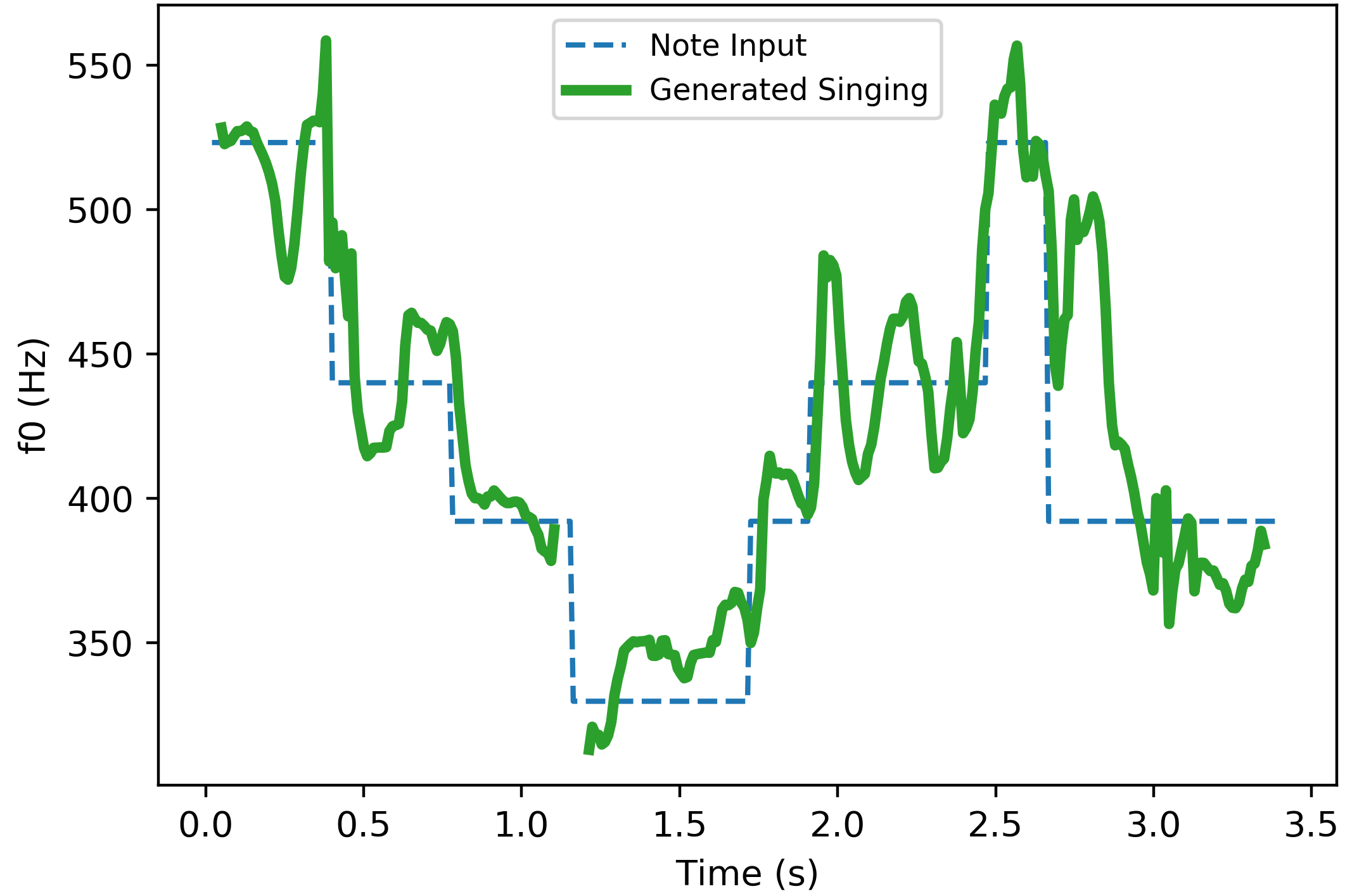}}
 \caption{generated Peking Opera singing $f_0$ contour by proposed system}
 \label{fig:f0_score_input}
\end{figure}

The generated $f_0$ contour from the synthesized singing is drawn with input music notes in Fig.~\ref{fig:f0_score_input}. Different from the pitch contour generated by normal TTS or singing systems, generated pitch contours by our proposed system in Peking Opera show more up and downs and with larger variation. The vibratos and transitions curves in Fig.~\ref{fig:f0_score_input} is consistent with Peking Opera singing style and shows the ability of proposed system to generate the expressiveness in Peking Opera.

% \begin{table}[]
% \caption{The Mean Opinion Score results of the proposed system and the original singing with 95\% confidence interval.}
% \label{mos-results}
% \begin{center}
% \begin{tabular}{@{}ccccc@{}}
% \toprule
%                   & Mean Opinion Score       \\ \midrule
% Original singing  & 4.57 $\pm$ 0.32    \\
% Proposed system  & 3.14 $\pm$ 0.35   \\ \bottomrule
% \end{tabular}
% \end{center}
% \vspace{-6mm}
% \end{table}

% \begin{figure}[bht] %figure*
%  \centerline{
%  \includegraphics[width=\columnwidth]{pics/f0_compare_recon-2.png}}
%  \caption{The comparison of the $f_0$ of original recording and sample synthesized by the proposed system.}
%  \label{fig:f0compare}
% \end{figure}

\section{Conclusions}
\label{sec:conclusion}

 Improvisation and expressiveness in Peking Opera singing makes it extremely difficult to synthesize this classical performing art. With proposed MDN-based phoneme duration generation with Lagrange Multiplier optimization, our system can generate more accurate phoneme duration compared to the Fitting Heuristic phoneme duration scaling method. Pseudo music notes are generated through the melody transcription algorithm to solve the score inconsistency problem in training. Both the objective average predicted phoneme duration error and the generated pitch contour show our system performances well in generating Peking Opera singing. And as one can see from MOS and the generated samples that there is still a gap between the generated singing and the real performance in terms of naturalness. Our further work includes collecting and labelling more Peking Opera singing data, conducting MOS test in larger scale with subjects in musical background, and improving the quality and pitch accuracy of the generated singing.

%\vfill\pagebreak

% References should be produced using the bibtex program from suitable
% BiBTeX files (here: strings, refs, manuals). The IEEEbib.bst bibliography
% style file from IEEE produces unsorted bibliography list.
% -------------------------------------------------------------------------
\nocite{black2014automatic}
\nocite{umbert2015expression}
% Generated by IEEEtran.bst, version: 1.13 (2008/09/30)

\bibliographystyle{IEEEtran}

\end{document}